\documentclass[twocolumn]{aastex62}

\submitjournal{ApJ}

\begin{document}

\title{Energetic gamma-ray emission from solar flares}

\correspondingauthor{Ervin Kafexhiu}
\email{ervin.kafexhiu@mpi-hd.mpg.de}

\author{Ervin Kafexhiu}
\affil{Max-Planck-Institut f\"ur Kernphysik, Saupfercheckweg 1, D-69117 Heidelberg, Germany}

\author{Carlo Romoli}
\affiliation{Max-Planck-Institut f\"ur Kernphysik, Saupfercheckweg 1, D-69117 Heidelberg, Germany}
\affiliation{Dublin Institute for Advanced Studies, 31 Fitzwilliam Place, Dublin 2, Ireland}

\author{Andrew~M.~Taylor}
\affiliation{DESY, D-15738 Zeuthen, Germany}

\author{Felix~Aharonian}
\affiliation{Max-Planck-Institut f\"ur Kernphysik, Saupfercheckweg 1, D-69117 Heidelberg, Germany}
\affiliation{Dublin Institute for Advanced Studies, 31 Fitzwilliam Place, Dublin 2, Ireland}

\begin{abstract}
Recent advances in the $\gamma$-ray observations of solar flares by the \textit{Fermi} satellite, demand revisions in the hadronic $\gamma$-ray flux computation below 1~GeV. In this work we utilize recently updated pion production cross sections, along with an accurate description of low energy nuclear interactions. Applying these new interaction descriptions to model the \textit{Fermi} Large Area Telescope (LAT) solar flare data, we infer the primary particle spectral parameters. Application of this new cross section description leads to significantly different spectral parameters compared to those obtained previously. Furthermore, the inclusion of nuclei in these calculations leads to a primary spectrum that is generally harder than that required from proton only considerations. Lastly, the flare data at lower MeV energies, detected by the \textit{Fermi} Gamma-ray Burst Monitor (GBM), is demonstrated to provide additional low-energy spectral information.
\end{abstract}


\section{Introduction \label{intro}}
Solar flares are powerful outburst phenomena observed in the solar atmosphere, recorded to release an energy of up to $10^{33}$ erg in short time ($10^2-10^3$ seconds) intervals (see e.g. \cite{Hudson1983, Kopp2005}). Their energy source is believed to be the magnetic energy stored in the solar corona, released through the magnetic reconnection. During these events, plasma heating up to keV temperatures, and ion (electron) acceleration up to energies of tens of GeV (hundreds of MeV) are observed. The largest solar flares are also associated with coronal mass ejection (CME); for a review see e.g. \cite{Aschwanden2002, Benz2008, Fletcher2011}. 

A significant fraction of the energy released during these events is inferred to be passed into non-thermal particles, with electrons dominating over protons at low ($<$~MeV) energies \cite{Aschwanden2017}. Despite their low-energy dominance, hard X-ray and $\gamma$-ray emission studies indicate that the spectra of electrons are soft above MeV energies (see e.g. \cite{Lin1982,Lin1985}). Furthermore, the $\gamma$-ray spectra of the brightest flares above 100~MeV would require extremely hard electron power-law spectrum ($J_e\sim E_e^{-\alpha}$ for $\alpha<2$)~\cite{Ajello2014} in order to be explained by an electron emission scenario. It is therefore natural to assume that hadrons dominate significantly over electrons at high energies, such that the $\gamma$-ray emission above 100~MeV is dominated by hadronic $\gamma$-ray production. 

Solar flare $\gamma$-ray emission up to 100~MeV was first detected from the GRS instrument on board of the \textit{Solar Maximum Mission} (SMM) \cite{Rieger1983}. Following this, the GAMMA-1 Telescope \cite{Akimov1991} and EGRET on board of the \textit{Compton Gamma-Ray Observatory} (CGRO) \cite{Kanbach1993}, detected $\gamma$-rays emission above 100~MeV, reaching energies up to 2~GeV. The launch of the \textit{Fermi} mission in 2008 started a new precision era in the study of high energy $\gamma$-rays from the sun. 

The \textit{Fermi} satellite has two detectors on board: the \textit{Fermi} Gamma-ray Burst Monitor (\textit{Fermi}-GBM) that is sensitive between 10~keV -- 30~MeV \cite{FermiGBM}, and the \textit{Fermi} Large Area Telescope (\textit{Fermi}-LAT) that is sensitive between 20~MeV -- 300~GeV \cite{FermiLAT}. The high statistics and energy resolution of the \textit{Fermi}-LAT has allowed accurate determination of the $\gamma$-ray spectra from solar flares. Moreover, recent release of the new PASS8 data has significantly increased the $\gamma$-ray sensitivity of the \textit{Fermi}-LAT below 1~GeV, of particular importance for the study of the ion distribution above 100~MeV/nuc in solar flare events.

Motivated by these recent improvements in solar flare observations, we here implement several improvements to the hadronic $\gamma$-ray production descriptions above 30~MeV. We firstly explore the application of new $p+p\to\pi$ cross sections, known to provide a particularly accurate description of the process for kinetic energies close to threshold. We also implement and explore the additional consideration of subthreshold pion and $\gamma$-ray continuum production above 30~MeV, produced via nuclear interactions, both of which have previously been neglected. 

The layout of this paper is the following. In Sec.~\ref{sec:Fermi} we consider the \textit{Fermi}-LAT data for four major solar flares and investigate the impact that the new PASS8 data has on two of these flares. In Sec.~\ref{sec:GammaChannels} we revise the $\gamma$-ray production cross sections and demonstrate explicitly the contributions of the subthreshold pions and the so called \textit{hard photon} channels, indicating the further impact that the consideration of different energetic particle abundances have on the final $\gamma$-ray spectrum. In Sec.~\ref{sec:ResultDiscuss} we discuss the primary spectra parameters, and conclude with a summary of the main results. 

\vspace{-0.3cm}

\section{\textit{Fermi}-LAT solar flare data \label{sec:Fermi}}

\subsection{\textit{Fermi}-LAT data}
We focus on four major solar flares, detected by \textit{Fermi}-LAT during solar cycle 24 between 2011 and 2013, for which $\gamma$-ray emission above an energy of 100~MeV was detected. They are: the flare 2011 March 7 and June 7 \cite{Ackermann2014}, the 2012 March 7 \cite{Ajello2014} and the 2013 October 11 \cite{Pesce-Rollins2015}, all analysed using the PASS7 \emph{Instrument Response Functions} (IRFs). The last two flares data are provided at different instances of their time evolution. These flares share a common feature, which is that their impulse phase is followed by a long and slowly varying $\gamma$-ray emission phase with $E_\gamma>100$~MeV. These data cover a wide energy interval from about 60~MeV to several GeV. All these spectral data carry similar features. They peak around $E_\gamma\approx200$~MeV and most of the data points above 1~GeV are upper limits.

\begin{table*}
\caption{Final values of the fit of the 2 solar flares that have been reanalyzed with PASS8 data. The parameter $\Phi_{100}$ indicates the flux above 100 MeV.}
\label{tab:2}
\begin{tabular}{ccccccc}
\toprule
Dataset         & \multicolumn{2}{c}{Power Law}&~~~~ & \multicolumn{3}{c}{Power Law+cut-off} \\
\cline{2-3} \cline{5-7}
                & $\Phi_{100}$ [$10^{-5}$ ph/cm$^2$/s] & $\Gamma$ & ~~~~& $\Phi_{100}$ [$10^{-5}$ ph/cm$^2$/s] & $\Gamma$ & $E_c$ [MeV] \\
\hline
2011 June 7      & $2.62 \pm 0.17$ & $2.45 \pm 0.07$ &~~~~& $3.22 \pm 0.21$ & $0.00 \pm 0.04$ & $103.6 \pm 6.6$ \\
2013 October 11 (07:16:40UT) & $14.9 \pm 0.4$  & $2.35 \pm 0.03$ &~~~~& $18.4 \pm 0.5$  & $0.13 \pm 0.17$ & $125 \pm 11$\\
2013 October 11 (07:35:00UT) & $22.7 \pm 0.7$  & $2.37 \pm 0.03$ &~~~~& $27.8 \pm 0.8$  & $0.22 \pm 0.17$ & $129 \pm 12$\\
\toprule
\end{tabular}
\end{table*}

\subsection{Reanalysis of the \textit{Fermi}-LAT data using the new PASS8 IRFs}
In 2015 \footnote{https://fermi.gsfc.nasa.gov/ssc/data/access/} the \textit{Fermi}-LAT Collaboration released a new IRFs version called PASS8. In comparison with the previously released software, there was a significant improvement in effective area, especially at energies below 1 GeV \footnote{ https://www.slac.stanford.edu/exp/glast/groups/ /canda/lat\_Performance.htm} which are particularly relevant for this study. For this reason, we asses the gain obtained with the reanalysis of the data focusing on two flares, namely the 2011 June 7 and the 2013 October 11. These two flaring events were chosen because they are short and easy to analyse without the need of particular techniques such as tracking \cite{Ackermann2014} or the use of tailored IRFs \cite{Ajello2014}.

The \textit{Fermi}-LAT data were analyzed with the standard binned likelihood method in an energy range from 60 MeV to 50 GeV. The region of interest (RoI) analyzed, was a square region of 24 degree size centred on the position of the Sun during the day of the flare. The localization of the centroid of the emission made use the data above 100 MeV to ensure a better point spread function, and was obtained using the standard tool {\tt gtfindsrc}. For the flare 2011 June 7, the Sun was close to the projected position of the Crab pulsar, so the centroid was extracted from a circle with a radius of 5 degrees to avoid contamination.

The emission of the Sun was modeled as a point like source centred in the centroid found previously and the model file used in the {\tt gtlike} routine included the diffuse model for the galactic and isotropic background as well. In the case of the 2011 flare, we added also the Crab pulsar. Because of the vicinity of this flare to the galactic plane, the normalizations of these extra sources were left free. For the 2013 flare the background models were instead fixed to the 3FGL catalogue \cite{2015ApJS..218...23A}.

The PASS8 data allows also an extra feature to reduce the amount of systematic uncertainties by taking into account the energy dispersion matrix. This step is particularly important when analysing, as in this case, energies below 100 MeV.

\begin{figure}
\includegraphics[scale=0.42]{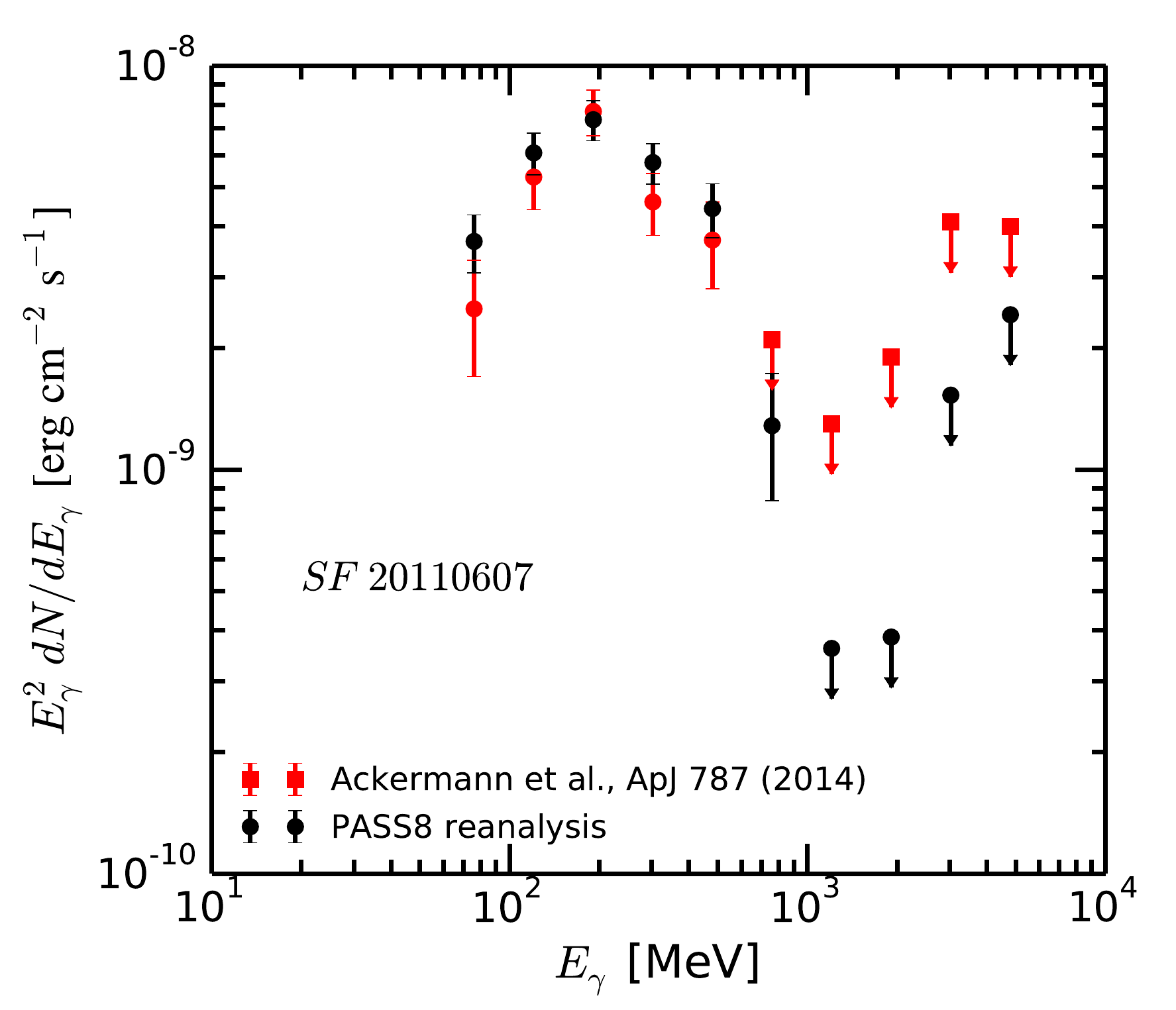}
\caption{Comparison between the \emph{Fermi}-LAT data of the solar flare 2011 June 7 reported in \cite{Ackermann2014} (red squares) and the reanalysis made using the PASS8 IRFs (black circles). Beside the reduction of the size of the error bars, the upper limits above 10 GeV are more constraining.\label{fig:comparisonP8}}
\end{figure}

The SED points were computed following the procedure illustrated in \cite{Ackermann2014}, fixing the power law index at 2 and leaving free the normalization in each energy bin. The SEDs can be seen in Figure \ref{fig:comparisonP8} for the 2011 flares shows the improvement in the determination of the spectrum using the new software.

In Table \ref{tab:2} the result of the likelihood fit on the re-analyzed data are shown using power-law and power-law with exponential cut-off functions. The comparison with the already published data shows significant differences only for the power law with exponential cut-off fit, having the cut-off energy reconstructed at slightly lower energies.

\begin{figure*}
\centering
\includegraphics[scale=0.42]{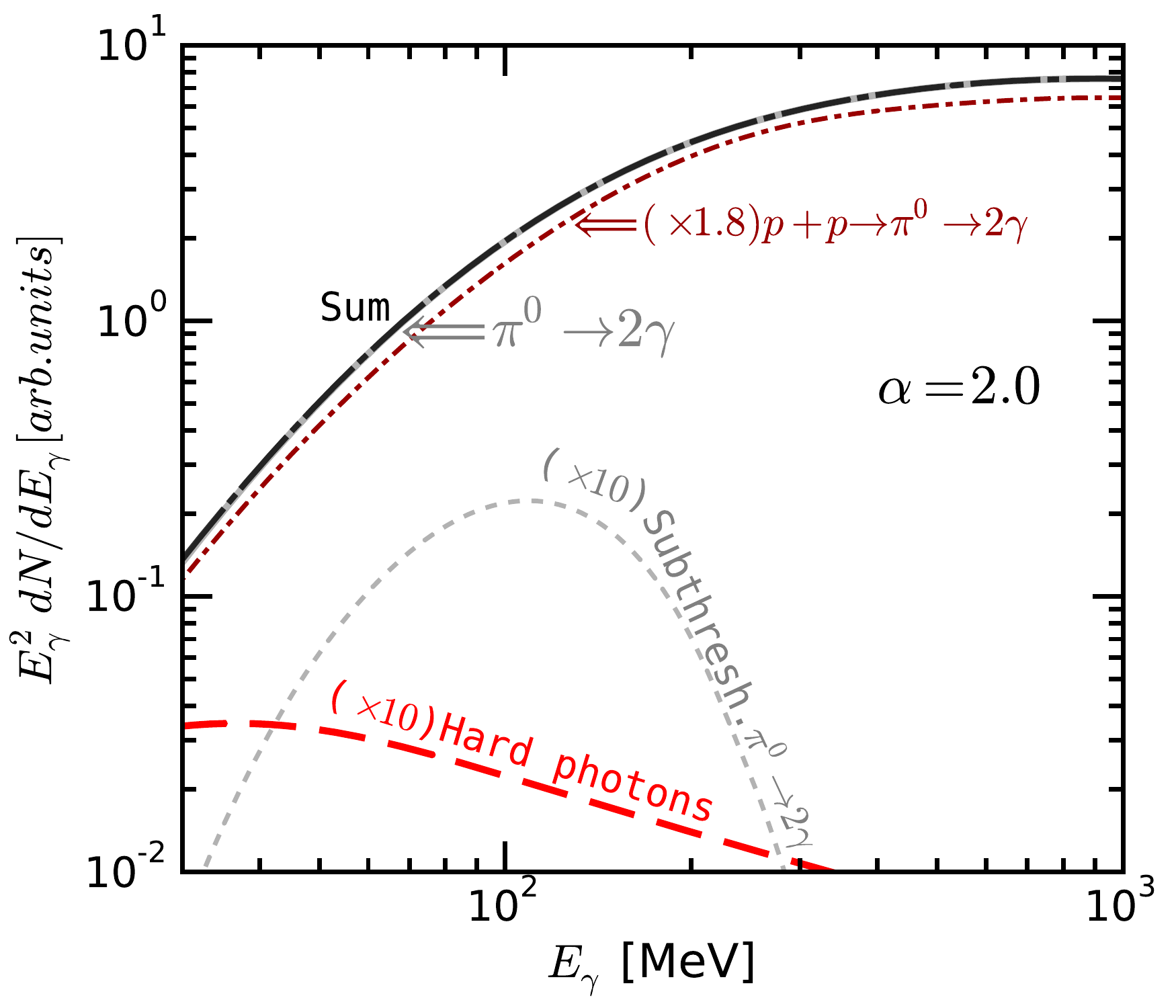}
\includegraphics[scale=0.42]{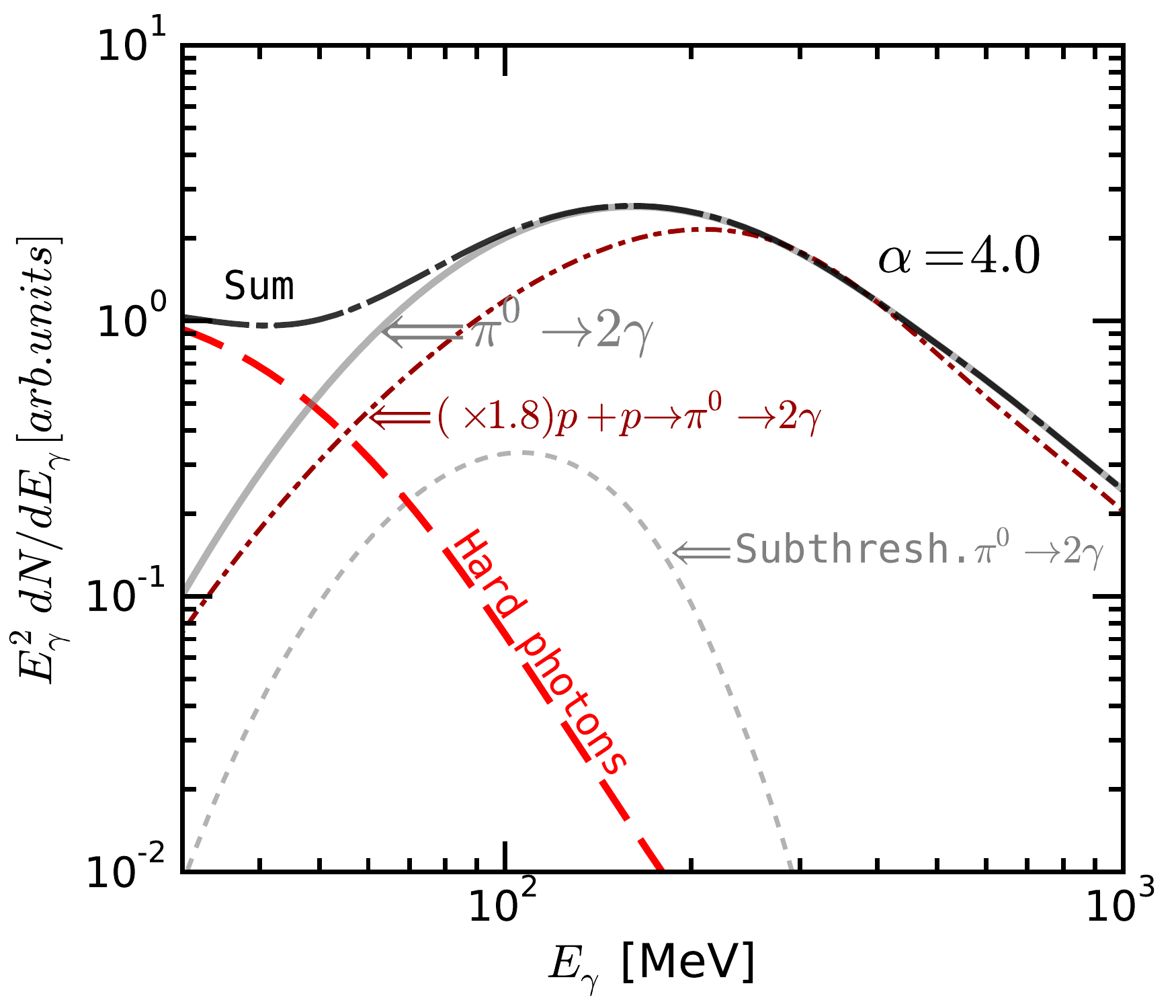}
\caption{Nuclear $\gamma$-ray spectrum for $E_\gamma>30$~MeV. The primary ion flux is a power-law function of kinetic energy per nucleon with $J_i= (f_i\times\upsilon) \sim T_i^{-\alpha}$ for $\alpha=2$ (left panel) and $\alpha=4$ (right panel), see Eq.~(\ref{eq:qs}). The mass composition of the projectiles and target material are set to SEP \cite{Reames2014Abund}, a solar composition \cite{Lodders2009}, respectively. The full gray line shows the $\pi^0\to2\gamma$ contribution, the gray dash line is only the contribution from the subthreshold pions. The red long dash line shows the contribution from \textit{hard photons}. The black long dash dot line is the sum of the \textit{hard photon} and $\pi^0\to2\gamma$ channels. For comparison the $p+p\to\pi^0\to2\gamma$ is also shown (brown short dash dot line) and is multiplied by 1.8 (the nuclear enhancement factor) to compare with the nuclear spectrum. In the left panel the contribution from the subthreshold pions and \textit{hard photons} is small and is multiplied by a factor 10 in the figure. \label{fig:Continuum}}
\end{figure*}

\begin{figure*}
\centering
\includegraphics[scale=0.5]{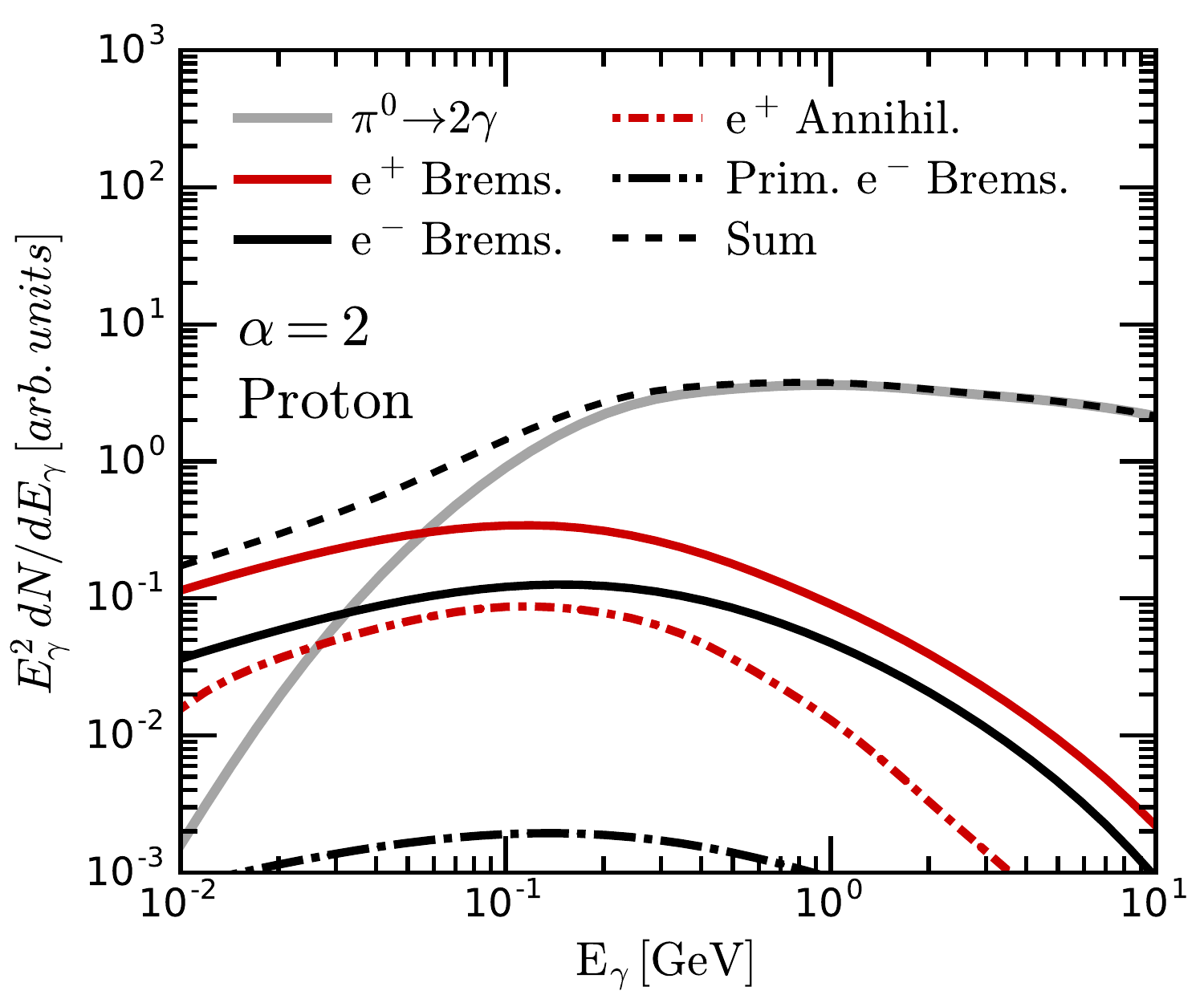}
\includegraphics[scale=0.5]{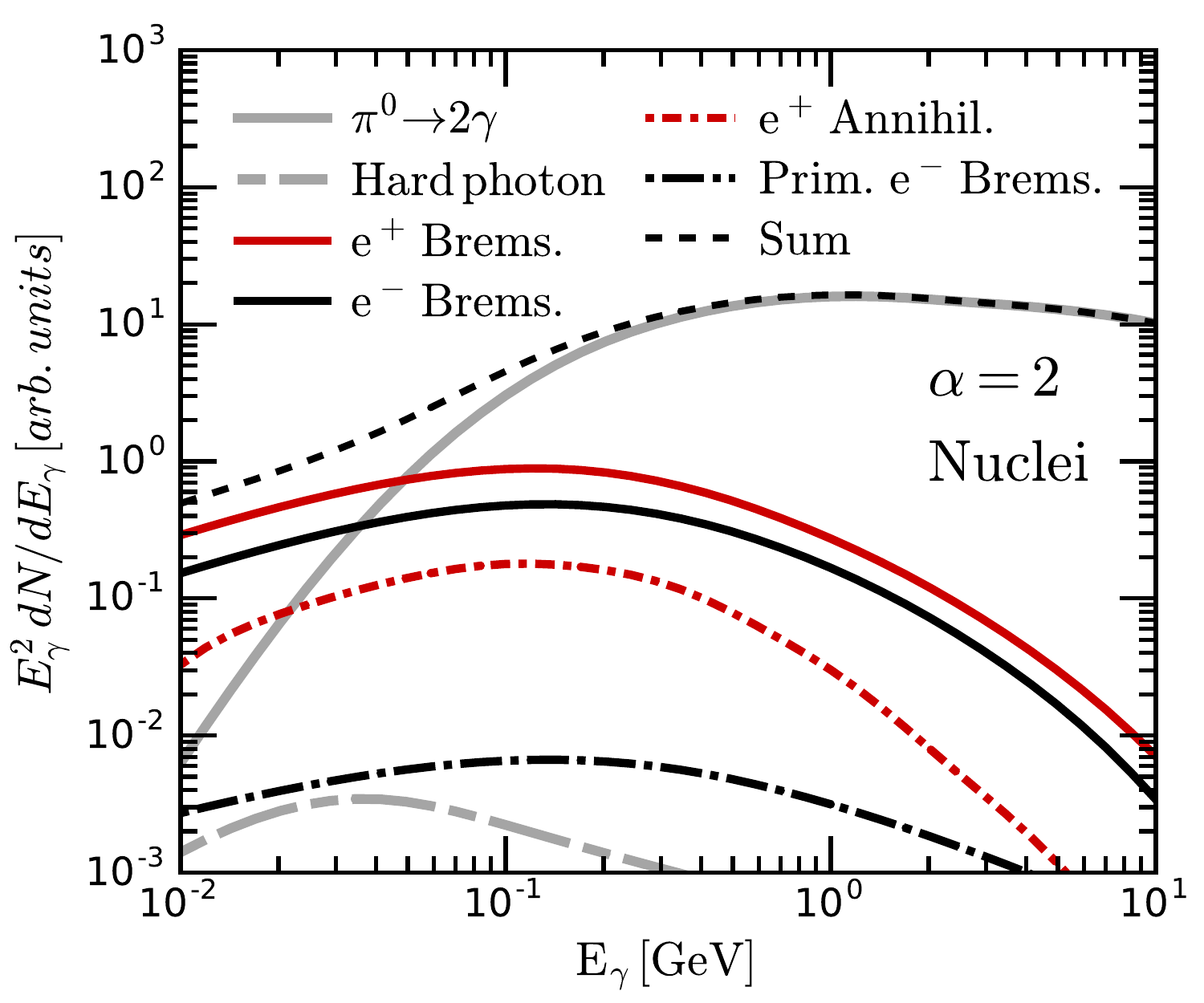}\\
\includegraphics[scale=0.5]{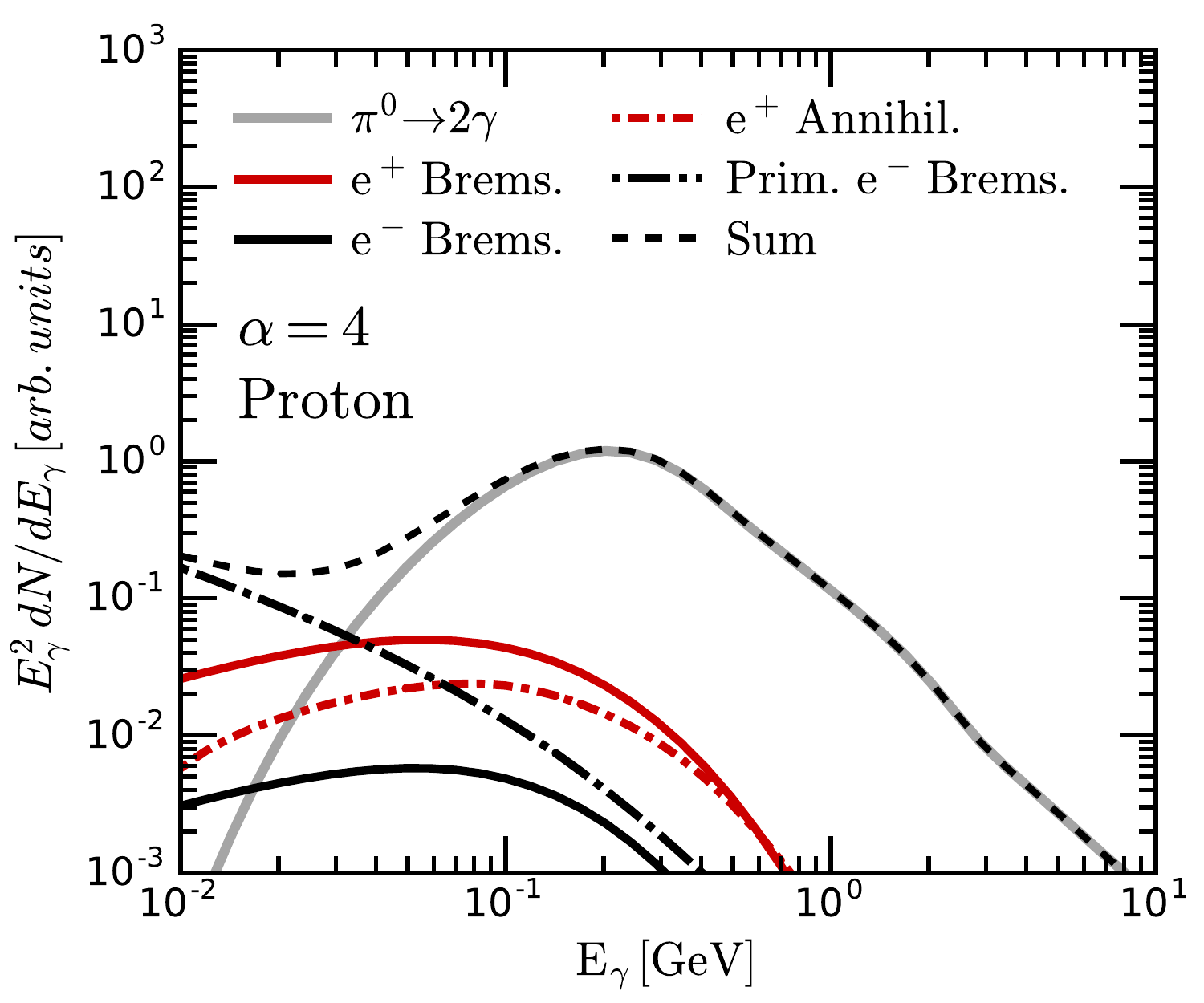}
\includegraphics[scale=0.5]{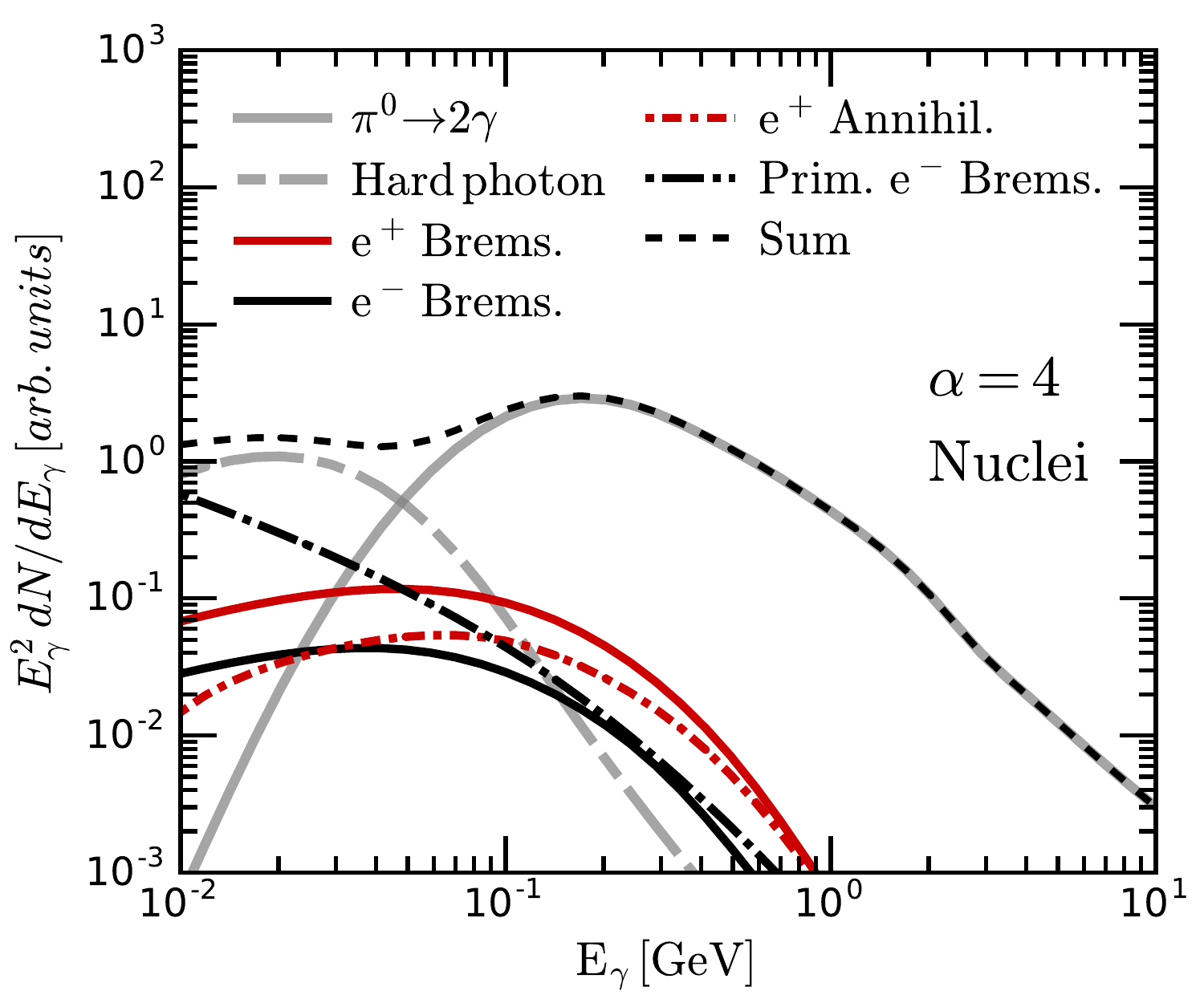}
\caption{Gamma-ray production spectrum from different leptonic and hadronic channels for $p+p$ interactions (left) and for SEP nuclei interacting with solar composition target (right). The primary ion flux considered, are power-laws in kinetic energy per nucleon $J_i \sim T_i^{-\alpha}$ for $\alpha=2$ (top panels) and $\alpha=4$ (bottom panels). The considered channels are: $\pi^0\to2\gamma$ (gray line), \textit{hard photons} (gray long-short dash line), $e^+$/$e^-$ bremsstrahlung (red/black line), $e^+$ annihilation in flight (red dash-dot line), primary electrons (black dash-dot line). The primary electron energy spectrum is assumed to be similar to proton one but with 1~\% of their flux. The black dash-line is the sum of all channels. \label{fig:1}}
\end{figure*}

\section{Gamma-ray production \label{sec:GammaChannels}}

\subsection{Interaction model}
We first consider a region in the solar atmosphere where accelerated primary particles interact with the ambient medium, producing secondary particles. Let $q_s(E_s,t)=d\dot{N}_s/dE_s$ be the secondary particle production rate per unit energy interval centered at energy $E_s$ at time $t$. The value of $q_s$ is computed as follows:
\begin{equation}\label{eq:qs}
q_s(E_s,t)= n_t \int\limits_{E^{\rm th}}^\infty dE\;f(E,t)\;\upsilon\; \frac{d\sigma}{dE_s}(E_i,E_s),
\end{equation} 

\noindent where, $n_t$ is the target medium number density, $E$ is the projectile energy, $E^{\rm th}$ is the threshold energy for the given reaction, $f(E,t)$ is the instantaneous energy distribution of the projectile particles in the interaction region, $\upsilon$ is the projectile speed and the $d\sigma/dE_s$ is the secondary particle production differential cross section for the specific process. It is clear from Eq.~(\ref{eq:qs}) that the computation of $q_s$ in a given target medium density $n_t$, requires the primary particle energy distribution $f$ and the specific process differential cross section and threshold energy.

Let us suppose that the energetic primary particles are injected in the interaction region with a rate per unit energy $Q$. Assuming that after being injected, these particles can only lose energy or escape from the interaction region, their $f$ evolves with time (see e.g. \cite{Ginzburg1964}):
\begin{equation}\label{eq:EqfpEvolv}
\frac{\partial f}{\partial t} + \frac{\partial}{\partial E} \left(\frac{E\, f}{\tau_{\rm Eloss}}\right) + \frac{f}{\tau_{\rm esc}} = Q.
\end{equation}

\noindent Here $\tau_{\rm Eloss}$ is the energy loss time and $\tau_{\rm esc}$ is the particle residence time in the interaction region. 

Note that the solution of Eq.~(\ref{eq:EqfpEvolv}) can be simplified for the two extreme limiting cases. In the first, the escape of particles from the region dominates over energy losses (ie. $\tau_{\rm esc}<\tau_{\rm Eloss}$), with the solution of Eq.~(\ref{eq:EqfpEvolv}) being $f=Q \times \tau_{\rm esc}$. In this case, the system is said to be in the \textit{thin target regime}. In the second case, if one can neglect particle escape (ie. $\tau_{\rm Eloss}<\tau_{\rm esc}$), the system is said to be in the \textit{thick target regime}. In this regime, $f$ evolves with time until the rate of injected particles in the region balances the rate of particles removed from it via energy losses. At this equilibrium point the evolution of $f$ reaches saturation.

\begin{figure*}
\centering
\includegraphics[scale=0.45]{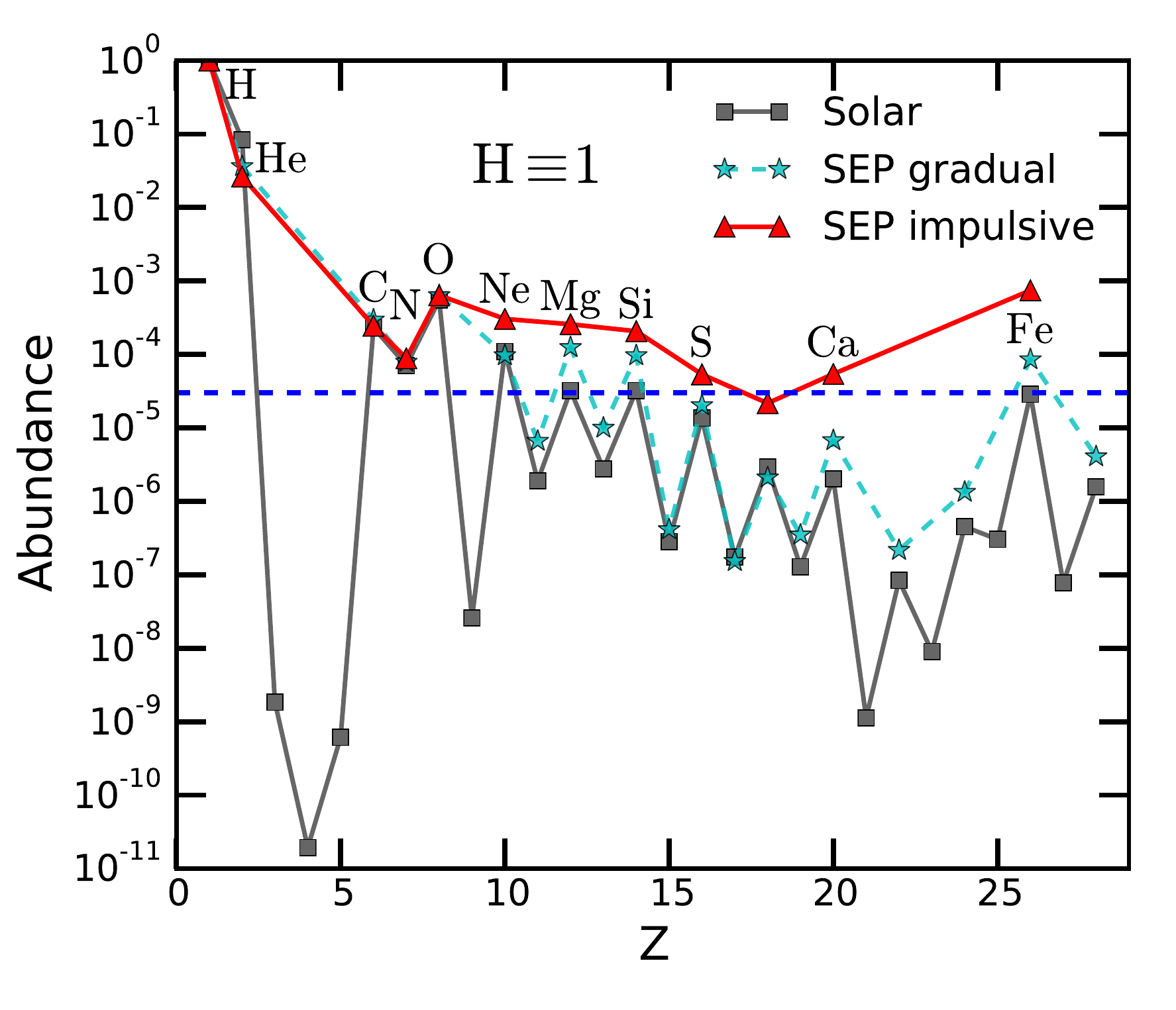}
\includegraphics[scale=0.45]{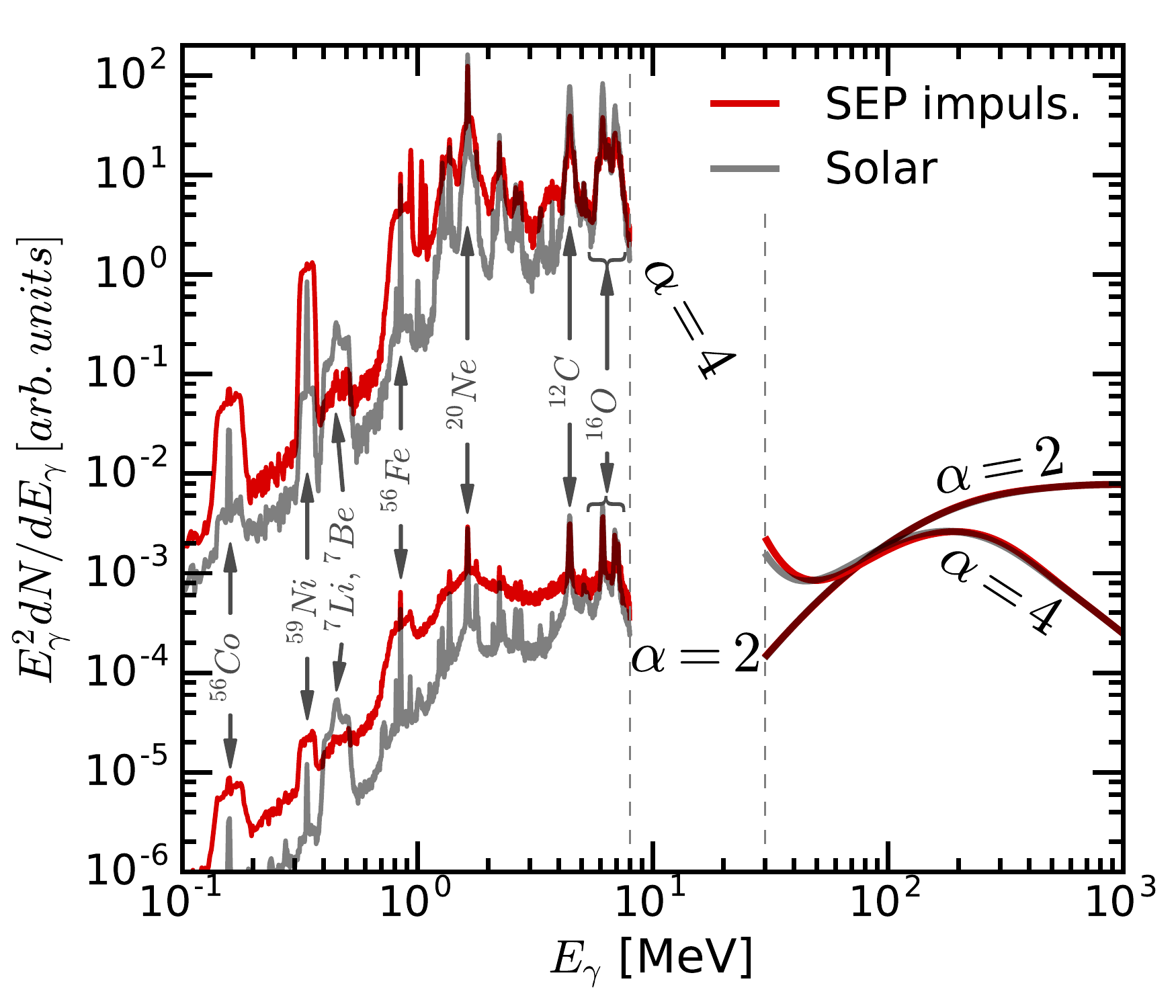}
\caption{Left panel shows the elemental abundances for a solar composition (gray squares) \cite{Lodders2009} and solar energetic particles (SEP) for gradual events (cyan stars) and impulsive events (red triangles) \cite{Reames2014Abund,Reames2014}. The blue dash-line shows the threshold we use in our calculations to select the elements. Right hand side panel shows the MeV and the GeV $\gamma$-ray spectra for two different compositions of energetic particles: impulsive SEP (red line) and solar composition (gray line). The energetic particle flux are power-laws in kinetic energy per nucleon with index $\alpha=2$ and 4. The final $\gamma$-ray spectra are normalized to have the same value at high energies. The elements that produce the strognest nuclear $\gamma$-ray lines are identified. The region between the vertical dash gray lines is dominated by the compound and preequilibrium nuclear $\gamma$-ray continuum that has not been taken into account and that smoothly connect the nuclear lines with the higher energy emission. \label{fig:CRSolar}}
\end{figure*}

\begin{figure*}
\centering
\includegraphics[scale=0.5]{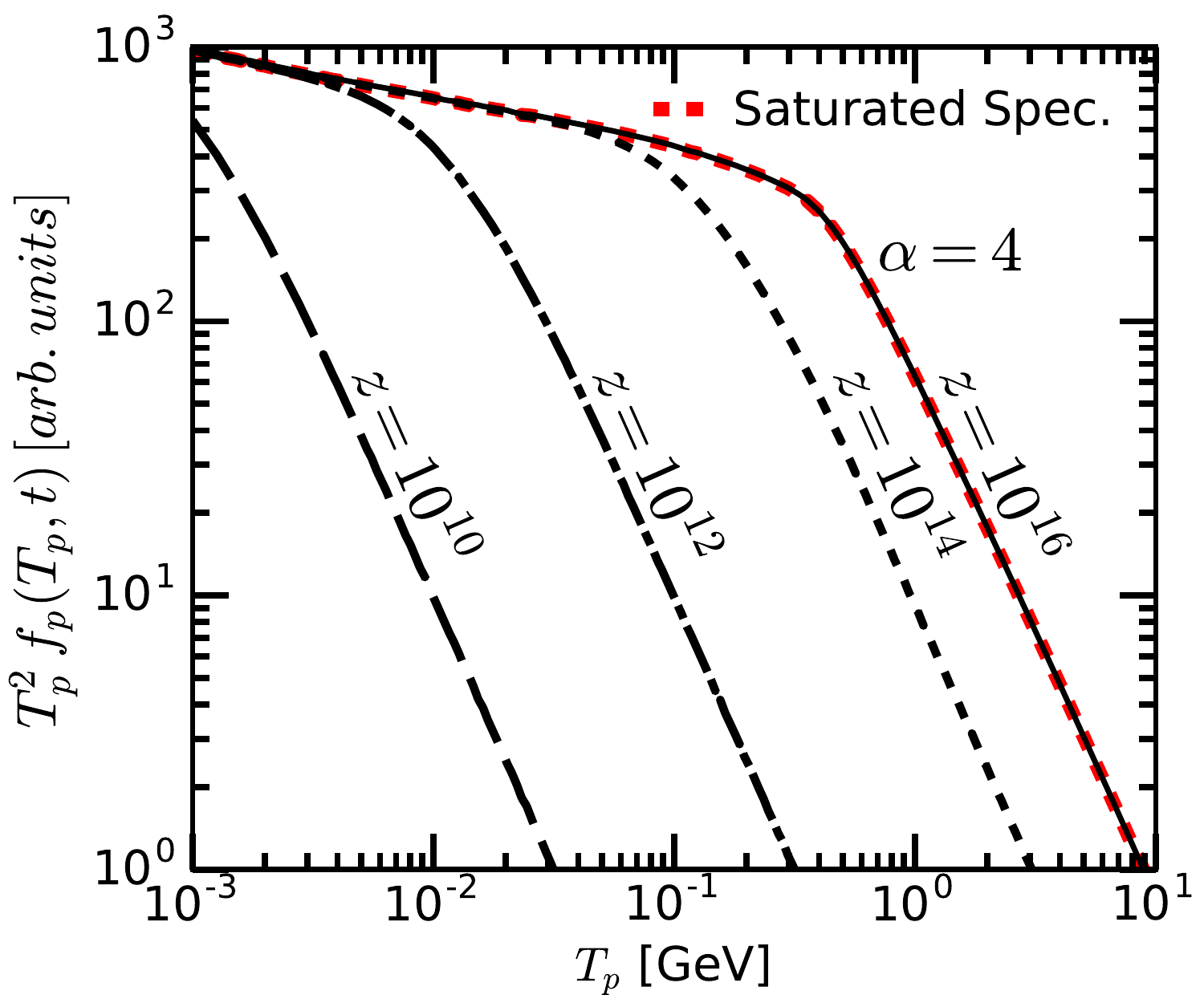}
\includegraphics[scale=0.5]{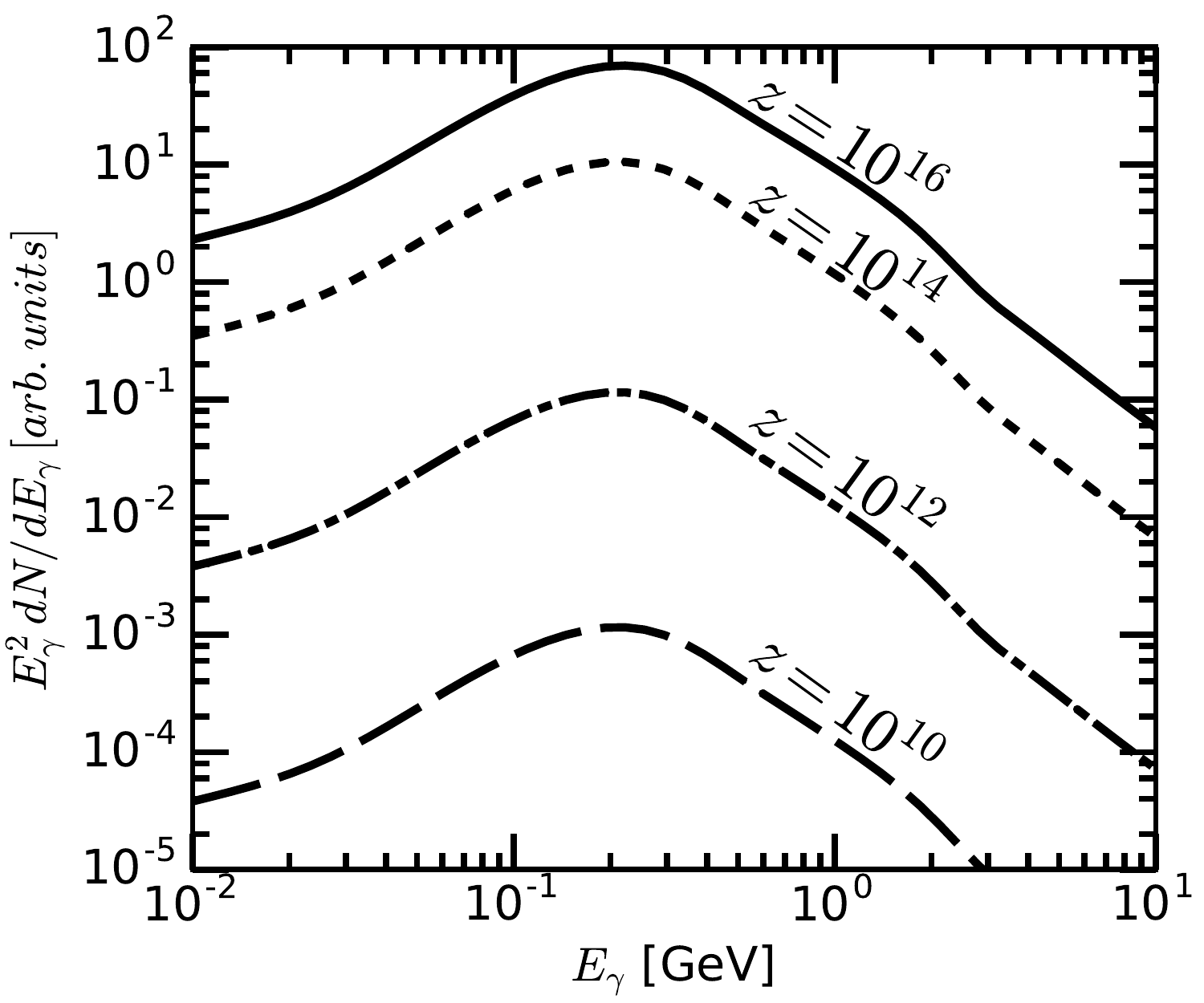}
\caption{Evolution of the proton energy distribution $f_p$ and the resulting $\gamma$-ray spectra from $p+p$ interactions. The proton injection rate is considered as a power-law function of the form $Q_p \sim T_p^{-\alpha}$ with $\alpha=4$. The number density and the magnetic field strength are set to $n_H=10^{13}\,{\rm cm^{-3}}$ and $B=100$~G, respectively. The left panel shows the proton energy distribution evolution for different values of the parameter $z=n_H\times t$ that are set to $z=10^{10}$ (long dash line), $10^{12}$ (dash dot line), $10^{14}$ (short dash line) and $10^{16}\,{\rm cm^{-3}\,s}$ (full line). For comparison, the saturated proton energy distribution is shown in red dash line which is reached for $z\gtrsim 5\times 10^{15}\,{\rm cm^{-3}\,s}$ ($t \gtrsim 5\times 10^{2}\,{\rm s}$). Their corresponding $\gamma$-ray spectra are shown on the right panel. \label{fig:evolution}}
\end{figure*}

\begin{figure}
\includegraphics[scale=0.42]{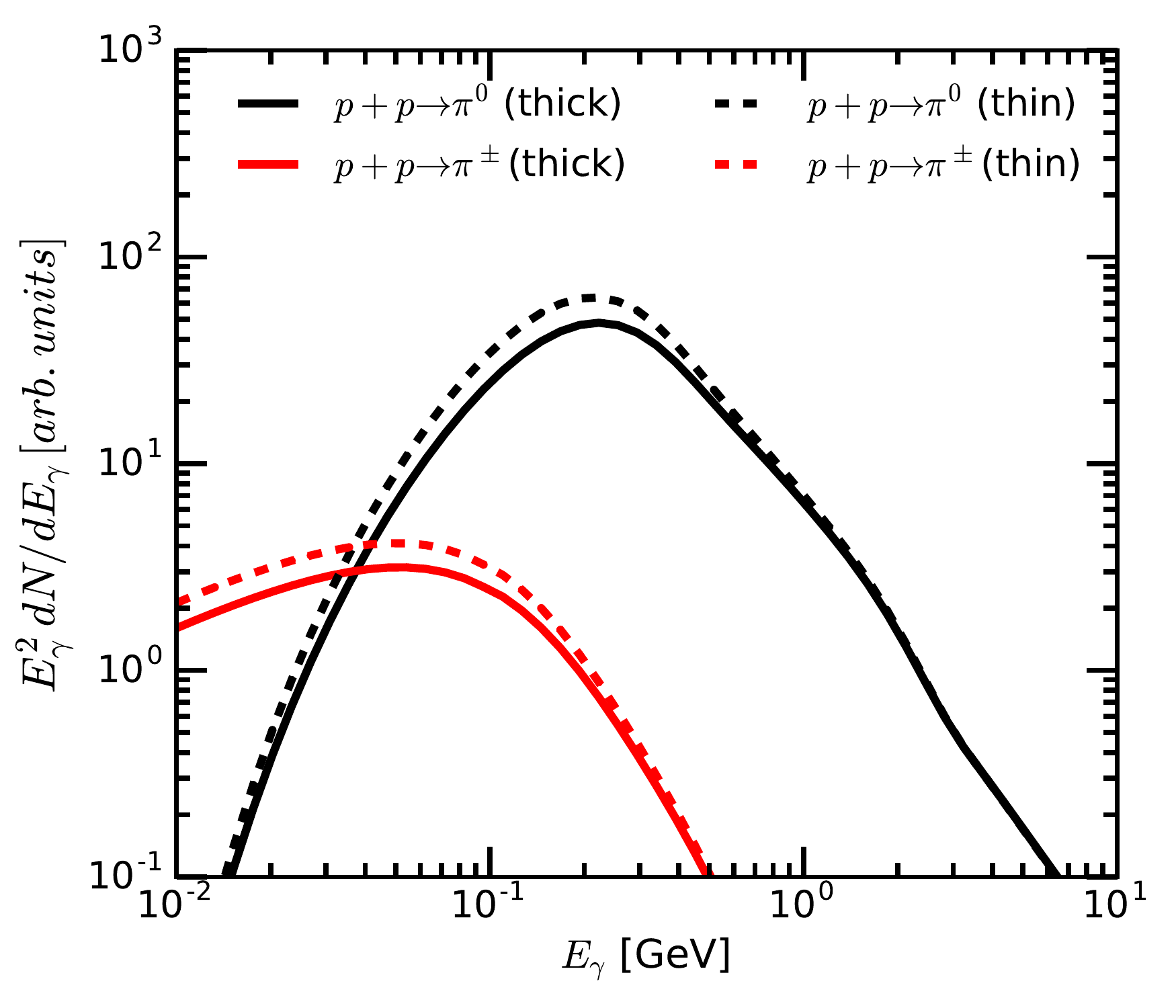}
\caption{ Gamma-ray spectrum from $p+p$ interactions for a thin target regime (dash line) and a saturated spectrum for a thick target regime (full line). The proton injection rate is assumed to be $Q_p \sim T_p^{-\alpha}$ for $\alpha=4$ and the number density and the magnetic field strength are set to $n_H=10^{13}\,{\rm cm^{-3}}$ and $B=100$~G. The black lines correspond to $\pi^0\to2\gamma$ decay and the red lines correspond to electron and positron bremsstrahlung and annihilation in flight assuming saturated $e^\pm$ spectra. \label{fig:thinthick}}
\end{figure}

\subsection{Gamma-ray production cross sections}

Solar flares can accelerate ions up to mildly relativistic energies. Accurate description of the $\gamma$-rays they produce thus requires accurate low energy $\gamma$-ray production cross sections. Hydrogen is the most abundant element in the solar atmosphere. Energetic protons with energies above the threshold energy of 0.28~GeV, colliding with the ambient hydrogen, can produce $\gamma$-rays through pion production interactions. Nuclei on the other hand, although less abundant than hydrogen, can also significantly contribute to the final $\gamma$-ray spectrum. Unlike proton interactions, nuclei have the advantage of being able to produce pions via the so-called \textit{subthreshold pion} production. Furthermore, they can also produce (below 0.28~GeV/nuc) direct continuum emission at energies $E_\gamma>30$~MeV via the so-called \textit{hard photon} production. These low energy $\gamma$-ray production channels can be especially important for solar flares, as these produce steep primary particle spectra with low energy cut-offs. These processes can significantly change the nuclear $\gamma$-ray spectral shape, compared to that produced for the simple proton-only case, and therefore should not be ignored.

The production cross-section for $p+p\to\pi^0\to2\gamma$ has been recently parametrized in \cite{Kafexhiu2014}. This new parametrization is particularly useful for solar flare modelling, achieving high accuracies down to the kinematic threshold. It uses recent pion production experimental data for kinetic energies $T_p<2$~GeV, and at higher collision energies utilizes a Monte Carlo description. We adopt here this parametrization to compute the $\gamma$ spectra from $p+p$ collisions.

The production cross sections for $p+p\to\pi^\pm$ at low collision energies near the kinematic threshold, has also been recently parametrized in \cite{PionBump}. In this work the charged pion energy distribution in the laboratory frame has been parametrized as a function of proton collision energy using the Geant4 toolkit \citep{Geant42003, Geant42006}. The total cross sections of the charged pion production, on the other hand, are parametrized using publicly available experimental data, see \cite{PionBump}. We adopt this parametrization to compute the electron/positron spectra from $p+p$ collisions.

The $\gamma$-ray production cross sections for low energy nuclei interactions, including the production of subthreshold pions and \textit{hard photons}, have recently been parameterized in \cite{Subthresh2016}. These parametrizations are based on publicly available experimental data and give simple and accurate analytical formulae that are valid for ion kinetic energies $T_i\leq100$~GeV/nuc. We adopt here these formulae to compute the $\gamma$-ray and the $e^\pm$ spectra from all possible nuclear interactions.

To compute the electron and positron spectra from the decay of charged pions, produced via $p+p$ and nuclear interactions, we convolve the $\pi^\pm$ spectra with the $e^\pm$ energy distribution function for the monoenergetic pions \citep[see e.g.][]{Scanlon1965,Dermer1986}. The electron and positron spectra are then used to compute the $\gamma$-ray spectrum from the bremsstrahlung  \cite{Blumenthal1970} and annihilation in flight \cite{Aharonian1981, Aharonian2000}. We note that in the case of nuclei, bremsstrahlung emission has a $Z^2$--dependence and the annihilation in flight has a $Z$--dependence on the nuclear charge number $Z$.

Although not the primary focus of this paper, nuclear interactions can also produce $\gamma$-ray emission below 30~MeV. The most prominent contributors of this emission are the nuclear $\gamma$-ray lines produced within the $0.1-10$~MeV energy interval, see e.g. \cite{Ramaty1979}. Their spectra have a strong dependence on the chemical composition of the target and projectile nuclei and the shape of the projectile particle spectrum below several hundreds of MeV/nuc. For the calculation of the nuclear $\gamma$-ray line spectra, we adopt the Monte Carlo method described in \cite{Ramaty1979, Kozlovsky2002}, describing the nuclear $\gamma$-ray lines below 8~MeV. Note that recent developments in both the experimental data and numerical descriptions have increased the accuracy of the nuclear $\gamma$-ray line spectra (see e.g. Refs.~\cite{Belhout2007, Murphy2009, Benhabiles-Mezhoud2011, Kiener2012}).

\begin{figure*}
\centering
\includegraphics[scale=0.45]{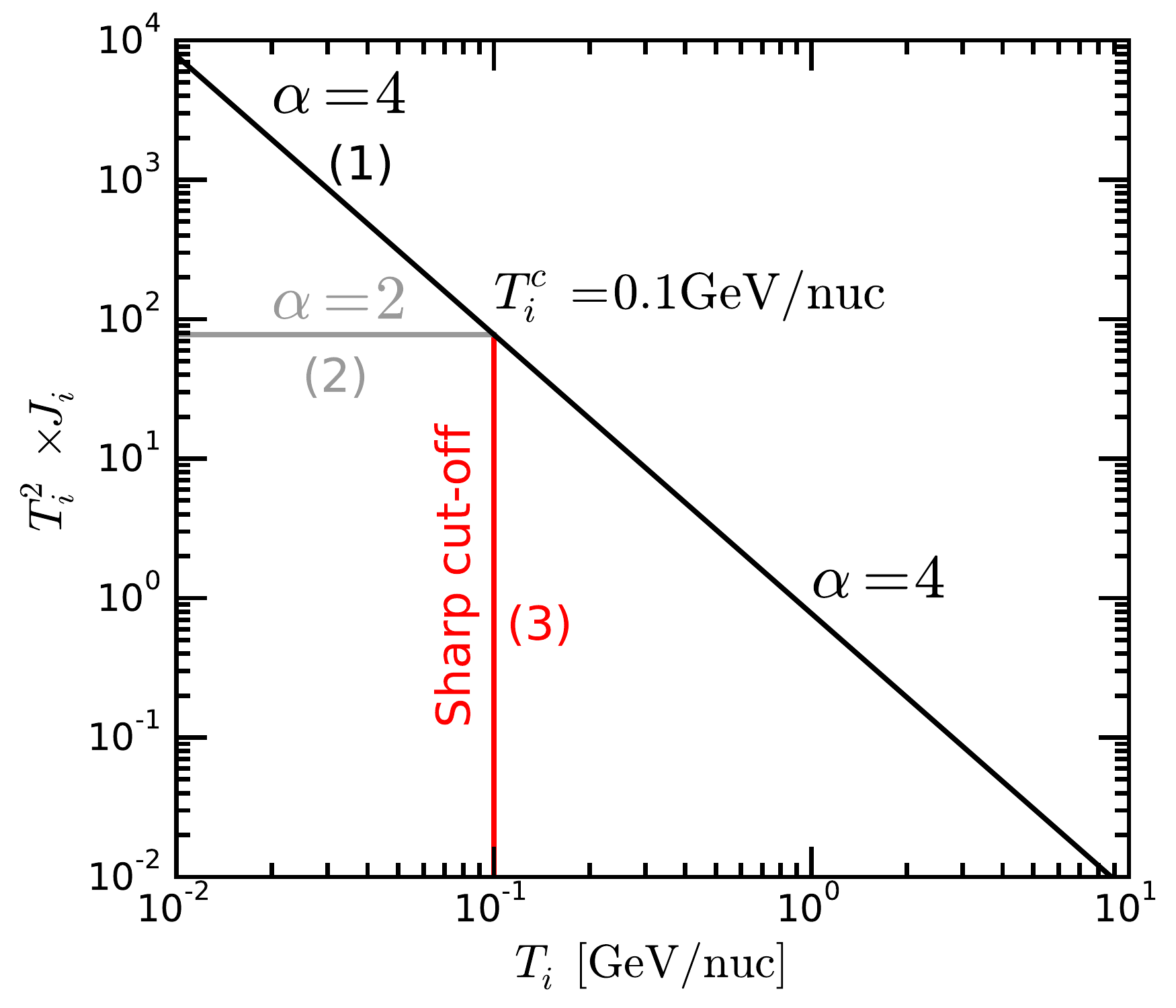}
\includegraphics[scale=0.45]{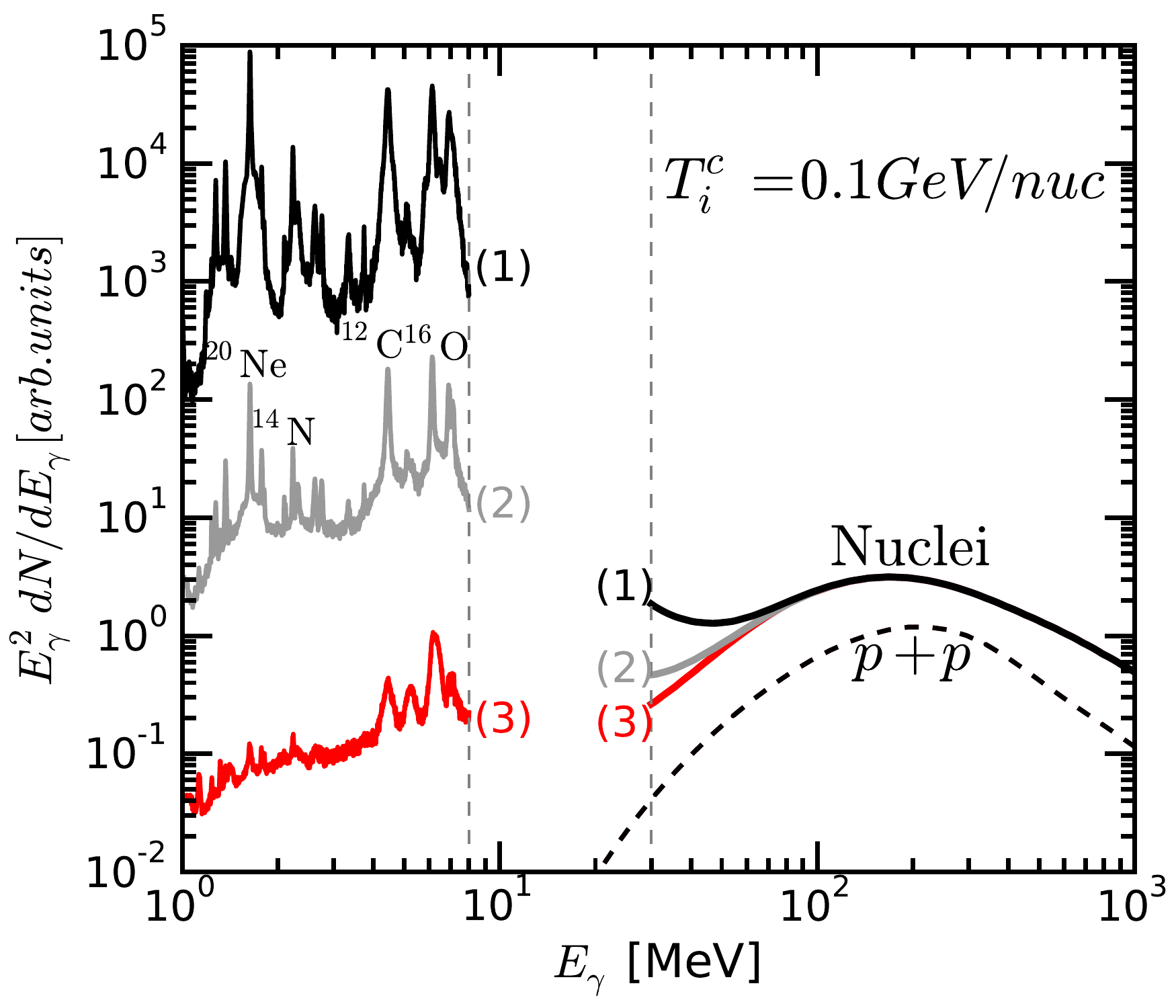}\\
\includegraphics[scale=0.45]{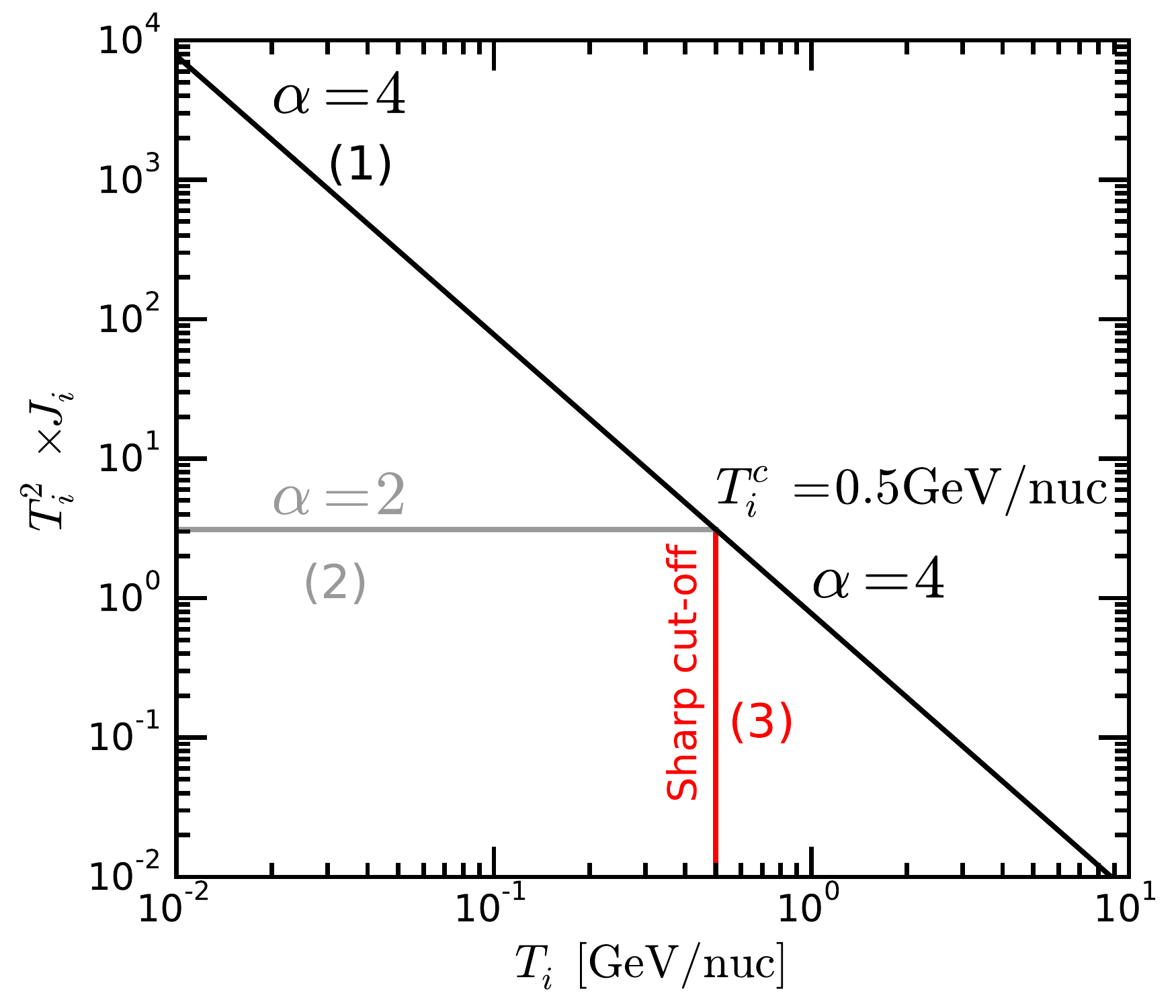}
\includegraphics[scale=0.45]{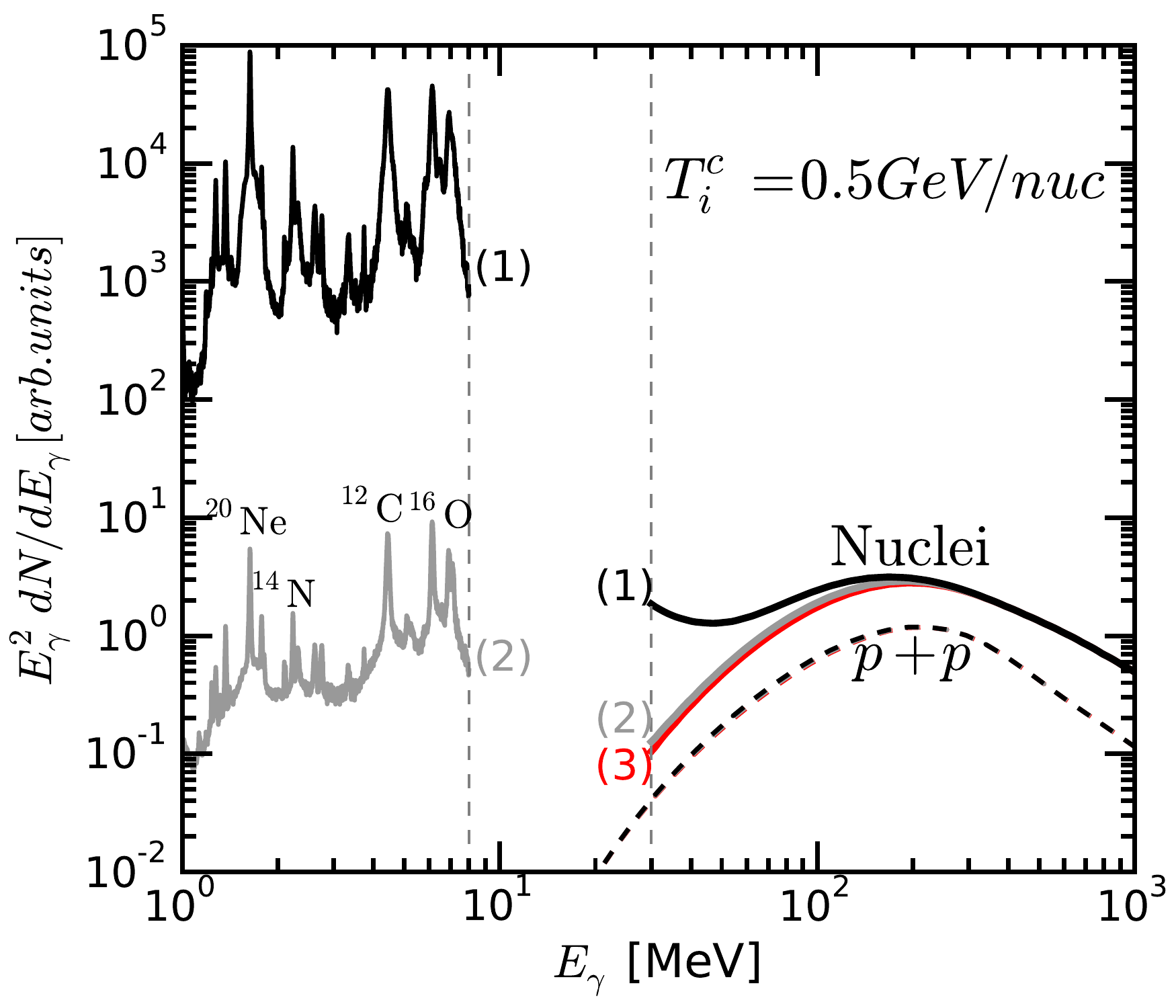}
\caption{Energy distribution of ions (left panels) and their corresponding broad-band $\gamma$-ray spectra (right panels). {\it Left panels}: Energy distribution of ions with power-law on the projectile kinetic energy per nucleon $T_i$ with index $\alpha=4$ that has a break at lower energies at $T_i^c=0.1$ (top) and 0.5~GeV/nuc (bottom). Curve (1) corresponds to the continuation of the power-law with $\alpha=4$ , curve (2) corresponds to $\alpha=2$ after the break and curve (3) implies a sharp cut-off at the break energy. {\it Right panels}: Corresponding $\gamma$-ray spectra due to: $\pi^0$ production; \textit{hard photons}; and nuclear $\gamma$-ray lines. The thin dash lines show the $p+p$ contribution. The region between the vertical dash gray lines is dominated by the compound and preequilibrium nuclear $\gamma$-ray continuum that has not been taken into account and that smoothly connect the nuclear lines with the higher energy emission. \label{fig:cuts2}}
\end{figure*}

\begin{figure}
\includegraphics[scale=0.45]{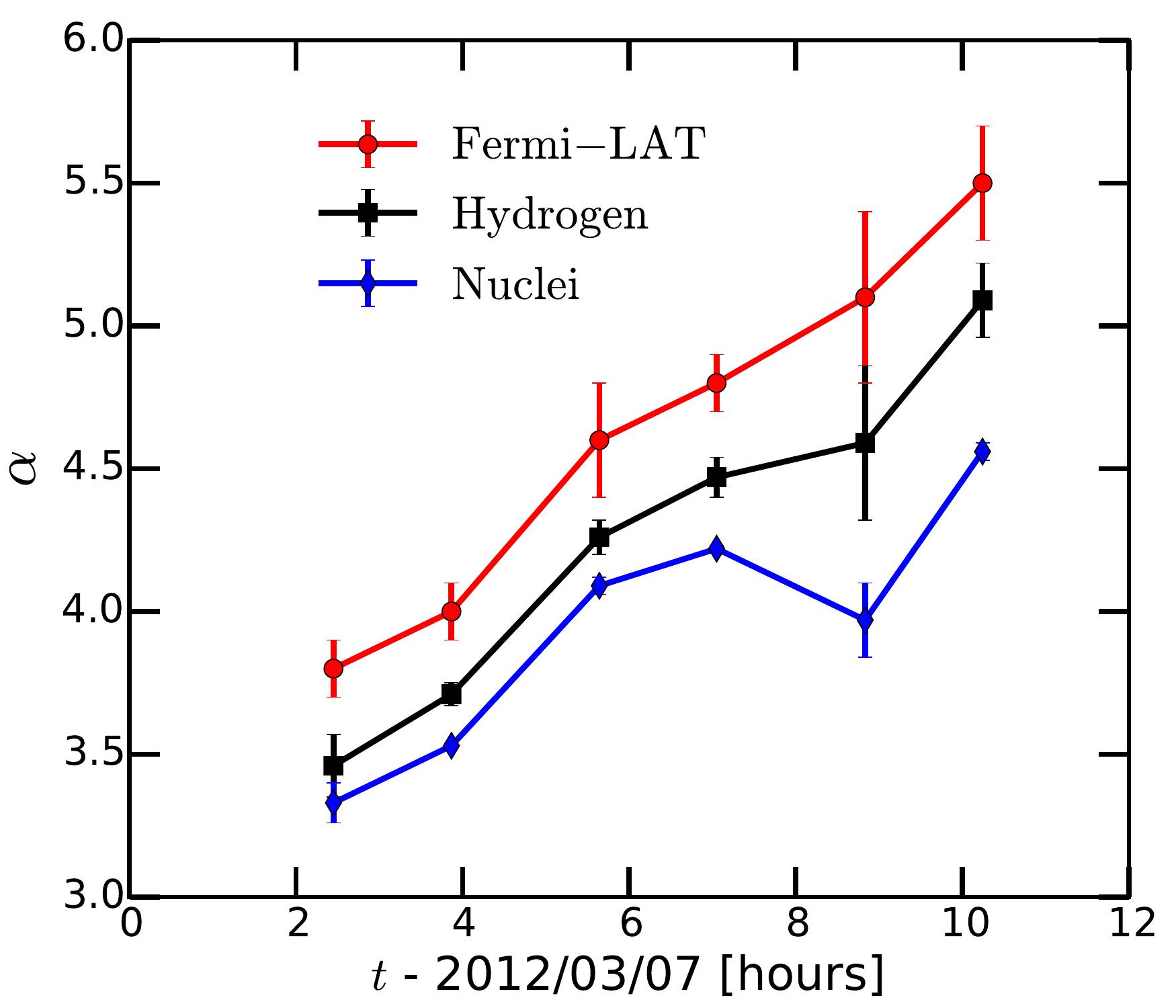}
\caption{Time evolution of the power-law index $\alpha$ for the solar flare 2012 March 7, see Table~\ref{tab:1}. The blue error bars show the results for nuclei (SEP interaction with solar composition target material), the black error bars show the results for pure hydrogen composition (using our updated $p+p$ cross sections \cite{Kafexhiu2014,PionBump}) and the red error bars show the results from \cite{Ajello2014}, see Table~\ref{tab:1}.  \label{fig:tab1}}
\end{figure}

\section{Gamma-ray emission \label{sec:illustrate}}

\subsection{Leptonic and hadronic channels}
We next apply the above discussed cross sections to the emission zone. As a first example we show the ensemble $\gamma$-ray emission from various channels following the interaction of energetic ions with the target gas. The flux of energetic ions is assumed to be a power-law function in kinetic energy per nucleon, $J_i=f_i\times \upsilon \sim T_i^{-\alpha}$. Indexes of $\alpha=2$ and $4$ are considered. The projectiles have a gradual solar energetic particle composition (gradual SEP) \cite{Reames2014Abund} and the target material has a solar composition \cite{Lodders2009}. We refer here to this abundance combination as ``Nuclei''. The resulting $\gamma$-ray spectra from both hadronic and leptonic emission are shown in Figs.~\ref{fig:Continuum} and \ref{fig:1}. In Fig.~\ref{fig:Continuum} we explicitly show the $\gamma$-ray emission from neutral pion decay including the subthreshold production, and \textit{hard photon} component of the hadronic emission. We note that the $\gamma$-ray spectrum for the proton-only interaction scenario is multiplied by 1.8 (the nuclear enhancement factor) to facilitate its comparison with the total nuclear spectrum. 

Figure~\ref{fig:1} includes the contribution from the leptonic $\gamma$-ray channels such as: $e^\pm$ bremsstrahlung, annihilation in flight and, primary electron bremsstrahlung. For this result we consider the maximum possible contribution from electrons, obtained for the case of a saturated (steady-state) $e^\pm$ spectra in the thick target regime. The $e^\pm$ injection rate is computed from the $\pi^\pm$ decays, whereas, for the primary electrons, we assume that their injection rate is similar to that of protons, but normalized to only 1~\% of the proton flux.

It is clear from Fig.~\ref{fig:1} that the presence of nuclei can have significant effects on the total $\gamma$-ray spectrum below $E_\gamma\lesssim200$~MeV. These effects are larger for the soft energetic particle spectral case considered ($\alpha=4$), for which proton-only interactions are unable to reproduce the $\gamma$-ray spectral shape. The additional inclusion of the leptonic channels further enhances such differences, see Fig.~\ref{fig:1}. For the example case shown here, the total differences between the proton and nuclear spectral shape are less than $60$~\% for $\alpha=2$ and a factor of 2 or more for the $\alpha=4$ case. It is therefore apparent that nuclear interactions produce notably different spectra compared to the proton-only case. Furthermore, the nuclear leptonic channels contribution is amplified by subthreshold $\pi^\pm$ production, a $e^-/e^+$ ratio close to unity, and the $Z^2$ dependence of the bremsstrahlung from the nucleus charge number $Z$. Note that unlike the low energy proton-only interactions where the ratio $e^-/e^+$ is close to zero, the low energy nuclear interactions produce a comparable amount of $e^\pm$ due to isospin symmetry and having equal number of protons and neutrons, see e.g. \cite{Subthresh2016}.

\subsection{Chemical composition of energetic particles}
We next explore the effect that different energetic particle chemical compositions have on the final $\gamma$-ray spectra. We adopt the same parameters as in the previous section, changing only the chemical composition of the energetic particles. Three of these compositions are considered: a solar composition (Solar), a gradual SEP and an impulsive SEP composition \cite{Reames2014Abund, Reames2014}. These abundances are plotted in the left panel of Fig.~\ref{fig:CRSolar}. As seen from the figure, the difference between a solar composition and a gradual SEP are not large. Therefore, when calculating the resulting $\gamma$-ray spectrum we consider only energetic particles with a solar or impulsive SEP type composition. The $\gamma$-ray spectra for these cases are shown on the right-hand panel of Fig.~\ref{fig:CRSolar}. 

In addition to the $\gamma$-ray continuum above 30~MeV we have also computed the spectrum of nuclear $\gamma$-ray lines below 8~MeV. It is clear from Fig.~\ref{fig:CRSolar} that changes in the mass composition from solar to impulsive SEP do not notably effect the $\gamma$-ray spectrum above 30~MeV, the differences in the range 30-200~MeV being less than 40~\%. However, the same changes in mass composition do have dramatic effects in the nuclear $\gamma$-ray line region. These differences originate from the fact that nuclei heavier than helium are more abundant for the impulsive SEP composition. Their excited emission subsequently suffering Doppler broadening effects. This results in the production of broad nuclear lines, which blend together to form a quasi-continuum for the SEP composition scenario.

\subsection{Proton thick target emission}
Due to energy losses, the proton energy distribution evolves with time in the interaction region, until reaching saturation (stready-state). Here we compute this evolution and the resulting $\gamma$-ray spectra. We adopt the thick target regime for protons, with a power-law injection spectrum of the form $Q_p=\mathcal{N}\times T_p^{-\alpha}$ with $\alpha=4$ and $\mathcal{N}$ a normalization constant. The electrons and positrons produced via $\pi^\pm$ production are also injected into the interaction region, and are also assumed to be in the thick target regime. Proton energy losses are dominated by ionization losses and inelastic collisions, whereas, the energy losses for electrons are dominated by ionization, bremsstrahlung and the synchrotron losses, see e.g. Refs.~\cite{Blumenthal1970, PDG2016}. Since the proton energy losses are proportional to the target number density $n_H$, their energy distribution is better described by the quantity $z=n_H\times t$. Here we assume that $n_H=10^{13}\,{\rm cm^{-3}}$ and a magnetic field strength $B=100$~G as fiducial values for the solar atmosphere.

The left panel of Fig.~\ref{fig:evolution} shows the proton energy distribution evolution at four distinguishable epochs with $z=n_H\times t=10^{10}$, $10^{12}$, $10^{14}$ and $10^{16}\,{\rm cm^{-3}\,s}$, corresponding to evolution times of $t=10^{-3}$, $10^{-1}$, $10^{1}$ and $10^{3}\,{\rm s}$, respectively. The proton energy distribution evolution can be understood in simple terms. When the evolution time is much smaller than the cooling timescale, the effect of losses is negligible. Therefore, the proton energy distribution has the same energy dependence as the injection rate $Q_p$, with the population of particles increasing linearly with time $f_p\sim Q_p\,t$; see e.g. $z=10^{10}\,{\rm cm^{-3}\,s}$ curve. However, when the evolution time $t$ becomes comparable with the cooling timescale, energy losses become important, shifting the high energy population of particles towards lower energies. Consequently, $f_p$ starts to deviate from $Q_p$, with $f_p$ eventually reaching its saturation shape, after which its evolution ceases.  For a steady injection rate, the proton energy distribution saturates for $z\gtrsim 5\times 10^{15}\,{\rm cm^{-3}\,s}$, corresponding for our example to $t \gtrsim 5\times 10^{2}\,{\rm s}$. Note that for an injection rate of the form $Q_p\sim T_p^{-\alpha}$, and energy losses of the form $\mathcal{P}\sim T_p ^\delta$, the saturated energy distribution is also a power-law $f_p\sim T_p^{-\beta}$ with $\beta=\alpha +\delta - 1$. This explains the broken power-law shape of the proton energy distribution in the non-relativistic region. The proton energy losses for $T_p>0.5$~GeV are dominated by inelastic collisions that have $\delta=1$, thus, $f_p$ has similar energy dependence as $Q_p$ because $\beta = \alpha$. At lower energies, however, where ionisation dominates energy losses, $\delta \approx -1$, the $f_p$ is a harder power-law with $\beta \approx \alpha -2$. The subsequent maximum value of the energy break is reached for the saturate spectrum with $T_p\sim0.4$~GeV, see Fig.~\ref{fig:evolution}.

The right-hand panel of Fig.~\ref{fig:evolution} shows the resulting $\gamma$-ray spectra evolution produced for the above described set-up, via proton-only interactions. We note that similar computations for nuclei are not straightforward. Due to nuclear reactions, the nucleus number of a given species changes in the interaction region. The presence of nuclear spallation processes cause evolution of the nuclear abundances that must also be taken into account when calculating the nuclear $\gamma$-ray spectrum. Such considerations, however, are beyond the scope of this paper.

Figure~\ref{fig:thinthick} shows the contribution of the leptonic and hadronic channels to the final $\gamma$-ray spectra for extreme cases, namely: the thin target regime and the thick target regime,  labelled ''thick'' and ''thin'' in the figure, respectively. Note that the radiation from $e^\pm$ bremsstrahlung and annihilation in flight are calculated for their saturated spectral cases, corresponding to their maximal potential contribution. We recall that the bremsstrahlung and annihilation in flight for a thin target $e^\pm$ regime can be negligible.

\begin{figure}
\includegraphics[scale=0.45]{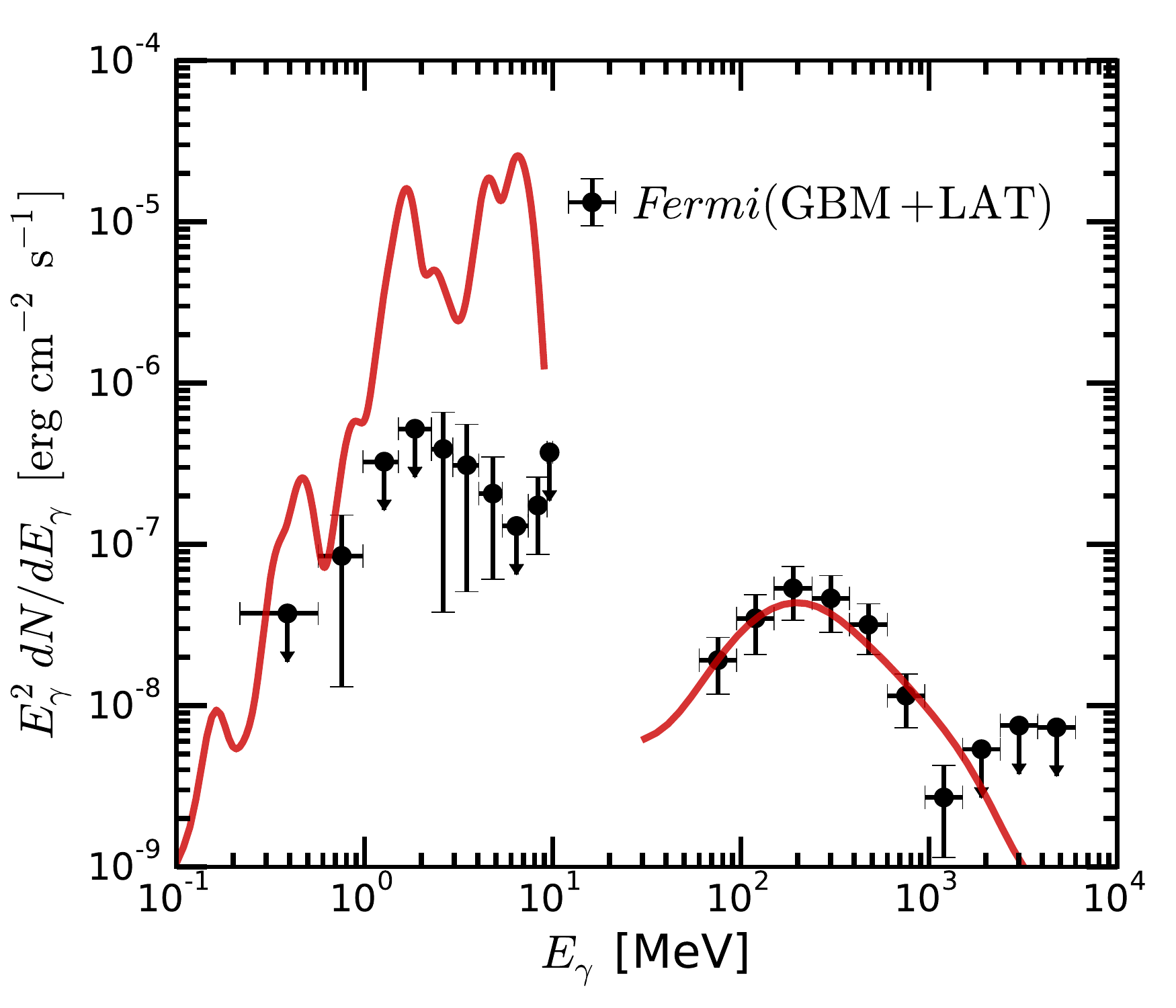}
\caption{The MeV and GeV $\gamma$-ray spectra from solar flare 2013 October 11 observed by the \textit{Fermi}-GBM and the \textit{Fermi}-LAT \cite{Pesce-Rollins2015}. The red line is the calculation using a power-law ion flux with index $\alpha=3.7$ derived by fitting the \textit{Fermi}-LAT data. The $\gamma$-ray flux  below 10 MeV is calculated assuming a continuation of the power-law function toward lower energies. The flux below 10 MeV is also smoothen to take into account the 10~\% energy resolution of the \textit{Fermi}-GBM detector. The energetic ion abundances are set to SEP and for the target material to a solar composition. \label{fig:lines_pi}}
\end{figure}

\subsection{Low energy spectra of energetic particles}
Here we explore the effect that the low energy spectral shape of the energetic particle spectra has on the $\gamma$-ray emission. We assume a thin target regime for simplicity, with the ion flux being described by a broken power-law. We consider break energies of $T_i^c=100$ and 500~MeV/nuc. The high energy part of the power-law has a fixed index of $\alpha=4$. The shape below the break energy is described by: 1) a continuation of the $\alpha=4$ power-law, 2) an $\alpha=2$ power-law and 3) a sharp low energy cut-off.

The left-hand panel of Fig.~\ref{fig:cuts2} shows the energetic particle spectra, whereas, the right-hand side shows the respective $\gamma$-ray spectra. We do not include here the $\gamma$-ray production from secondary $e^\pm$ channels. It is clear from the figure that the low energy primary spectral shape has a dramatic effect on the $\gamma$-ray spectrum below 200~MeV, especially in the energy region of the nuclear $\gamma$-ray lines, where the emissivities can change by orders of magnitude. These effects will be magnified if the solar composition is replaced by a heavier one. Unlike the nuclear interactions, the resulting radiation spectrum from proton-only interactions remains practically unchanged.

\begin{table}
\caption{The MCMC results for the primary spectrum power-law index $\alpha$ for the considered \textit{Fermi}-LAT solar flare data. The ``Hydrogen'' are the results for pure hydrogen compositon, whereas, ``Nuclei'' are the results for the SEP interacting with a solar composition target material. The ``\textit{Fermi}'' column quotes the index $\alpha$ values that are published in \textit{Fermi}-LAT publications \cite{Ackermann2014, Ajello2014, Pesce-Rollins2015}. \label{tab:1}}
\begin{tabular}{lccc}
\toprule
 Flare  & Hydrogen & Nuclei & \textit{Fermi} \\
\hline
2011 March 7                 & $4.27^{+0.22}_{-0.20}$ & $3.80^{+0.11}_{-0.09}$ & $4.5^{+0.2}_{-0.2}$ \\
2011 June 7                  & $4.12^{+0.54}_{-0.43}$ & $3.48^{+0.19}_{-0.14}$ & $4.3^{+0.3}_{-0.3}$ \\
2012 March 7 (02:27:00UT)    & $3.46^{+0.13}_{-0.11}$ & $3.33^{+0.09}_{-0.07}$ & $3.8^{+0.1}_{-0.1}$ \\
2012 March 7 (03:52:00UT)    & $3.71^{+0.04}_{-0.04}$ & $3.53^{+0.02}_{-0.01}$ & $4.0^{+0.1}_{-0.1}$ \\
2012 March 7 (05:38:32UT)    & $4.26^{+0.10}_{-0.06}$ & $4.09^{+0.04}_{-0.03}$ & $4.6^{+0.2}_{-0.2}$ \\
2012 March 7 (07:03:00UT)    & $4.47^{+0.07}_{-0.07}$ & $4.22^{+0.01}_{-0.01}$ & $4.8^{+0.1}_{-0.1}$ \\
2012 March 7 (08:50:00UT)    & $4.59^{+0.31}_{-0.27}$ & $3.97^{+0.14}_{-0.13}$ & $5.1^{+0.3}_{-0.3}$ \\
2012 March 7 (10:14:32UT)    & $5.09^{+0.21}_{-0.13}$ & $4.56^{+0.03}_{-0.03}$ & $5.5^{+0.2}_{-0.2}$ \\
2013 October 11 (07:16:40UT) & $3.98^{+0.30}_{-0.24}$ & $3.71^{+0.21}_{-0.20}$ & $3.8^{+0.2}_{-0.2}$ \\
2013 October 11 (07:35:00UT) & $3.88^{+0.26}_{-0.22}$ & $3.62^{+0.19}_{-0.18}$ & $3.7^{+0.2}_{-0.2}$ \\
\toprule
\end{tabular}
\end{table}

\section{Results and Discussion \label{sec:ResultDiscuss}}

In this section we show the energetic particle spectral parameters obtained by fitting the \textit{Fermi}-LAT solar flare data described in section~\ref{sec:Fermi}. For this analysis we assume a thin target regime for ions and a thick target regime for the secondary electrons (i.e. adopting their saturated spectra). We also consider two chemical compositions, namely: a pure proton (hydrogen) and an SEP composition (Nuclei), interacting with solar abundance target material. We note that changing the chemical composition of energetic particles from gradual to impulsive SEP or to a solar composition has negligible effects in the energy range $E_\gamma\geq60$~MeV relevant for the \textit{Fermi}-LAT solar flare data, see Fig.~\ref{fig:CRSolar}. We consider a primary ion flux described by a power-law function of the form $J_i=N\times T_i^{-\alpha}$, where, the normalization constant $N$ and the power-law index $\alpha$ are free parameters. For exploring this spectral parameter space, we adopt the Goodman and Weare's affine invariant Markov Chain Monte Carlo Ensemble sampler (MCMC) as is implemented in \cite{emcee} and adopt the revised $\gamma$-ray production cross sections described in section~\ref{sec:GammaChannels}. The results of the analysis are summarized in Table~\ref{tab:1}.

We next compare the results obtained here for the hydrogen case with the same ones quoted in the \textit{Fermi}-LAT publications. As seen in Table~\ref{tab:1}, the index $\alpha$ obtained in this work has significant deviations from the values quoted in the literature. These changes can be predominantly explained by the differences in the $p+p$ cross sections adopted between our revised parametrizations and the ones used in the \textit{Fermi}-LAT publications \cite{Dermer1986a,Murphy1987}.

Further significant differences are also seen when nuclei are considered. The index $\alpha$ required to fit the $\gamma$-ray data is systematically smaller for nuclei than for the hydrogen case, see Table~\ref{tab:1}. Thus, for nuclei, the same $\gamma$-ray data require a harder primary spectrum than the corresponding proton-only values. These contrasts in the primary particle parameter space are a reflection of their different $\gamma$-ray spectral shape for $E_\gamma<200$~MeV.

Observations of the 2012 March 7 and 2013 October 11 solar flares by \textit{Fermi}-LAT has provided $\gamma$-ray data at different instances during the evolution of the flares. Specifically, the analysis of the 2012 March 7 flare data suggests that the power-law index $\alpha$ increases with time, see Fig.~\ref{fig:tab1}.

For the 2013 October 11 flare, the \textit{Fermi}-GBM data below 10~MeV are also provided \cite{Pesce-Rollins2015}. Figure~\ref{fig:lines_pi} shows the subsequent best-fit $\gamma$-ray spectrum to the \textit{Fermi}-LAT data, with a low energy comparison to the \textit{Fermi}-GBM data for the Nuclei composition case. We assume here that the same functional form of the primary spectra fit to the \textit{Fermi}-LAT data continues down to the lower energies relevant for nuclear $\gamma$-ray line production. As we can see in Fig.~\ref{fig:lines_pi}, the $\gamma$-ray flux predicted from the soft pure power-law primary flux fits well the high energy data, but over-predicts the MeV $\gamma$-ray flux. Note, however, that in the thick target regime, ionization losses will harden the non-relativistic part of the ion spectrum. The MeV $\gamma$-ray fluxes predicted here may therefore be reduced; e.g. see Fig.~\ref{fig:cuts2}. Furthermore, for the Nuclei composition case, with the energetic particles interacting in the thick target regime, the evolution of the nuclear states due to spallation will further complicate this picture. Interestingly, such an evolution may offer the future opportunity to probe the nuclear residence times using the nuclear $\gamma$-ray line information.

Lastly, we recall that our reanalysis of the 2011 June 7 and the 2013 October 11 data using the new PASS8 data shows improvements on the quality of the data by reducing the errorbars and by adding one more data point around 1~GeV, see Fig.~\ref{fig:comparisonP8}. Despite this, the final primary spectra parameters required to fit the $\gamma$-ray data do not show significant changes.

\section{Conclusions}

The high quality $\gamma$-ray solar flare observations carried out by \textit{Fermi}-LAT data, demands accurate modelling of this $\gamma$-ray emission above 30~MeV. In this work we have revised hadronic $\gamma$-ray emission calculations for both protons and nuclei, taking into account the secondary electrons produced. Utilizing our recent updates to the hadronic $\gamma$-ray production cross-sections for both protons and nuclei, the importance of the description of pion production close to threshold, nuclear subthreshold pion production, and \textit{hard photon} emission are highlighted. The neglection of these processes is found to be considerably detrimental in the recovery of the underlying projectile particle spectrum using the \textit{Fermi}-LAT $\gamma$-ray data.


\bibliographystyle{aasjournal}
\bibliography{refs}

\end{document}